%% file: main.tex
\documentclass{llncs}
\frontmatter          % for the preliminaries
\mainmatter              % start of the contributions
\title{MLSEB: Edge Bundling using Moving Least Squares
Approximation}

\titlerunning{Moving Least Squares Edge Bundling}
\author{Jieting Wu, Jianping Zeng, Feiyu Zhu and Hongfeng Yu}
\authorrunning{Jieting Wu et al.} % abbreviated author list (for running head)
\institute{University of Nebraska-Lincoln, Lincoln NE 68588, USA,\\
\email{\{jwu, jizeng, fzhu, hfyu\}@cse.unl.edu},
}

\usepackage{amsmath}
\usepackage{amssymb}
\usepackage{booktabs}
\usepackage{mathtools}
\usepackage{multirow}
\usepackage{floatrow}

\usepackage{graphicx}
\usepackage{colortbl,array} % für farbige cells
\usepackage{bigdelim}

\usepackage{wrapfig}

\newcommand{\R}{\mathbb{R}}

\newcolumntype{N}{>{\centering\arraybackslash}m{.5in}}
\newcolumntype{G}{>{\centering\arraybackslash}m{2in}}

\begin{document}

\graphicspath{{./}}

\maketitle              % typeset the title of the contribution

\begin{abstract}
\input{./abstract.tex}
\keywords{edge bundling, graph visualization, moving least squares, visualization quality}
\end{abstract}
%

%------------------------------------------------------------
\input{./sec_intro.tex}
%------------------------------------------------------------
\input{./sec_related.tex}
%------------------------------------------------------------
% \input{../section/sec_background.tex}
%------------------------------------------------------------
\input{./sec_method.tex}
%------------------------------------------------------------
\input{./sec_implementation.tex}
%------------------------------------------------------------
\input{./sec_result.tex}

%------------------------------------------------------------
%\input{../section/sec_assess.tex}
%------------------------------------------------------------
\input{./sec_conclusion.tex}

\section*{Acknowledgment}

This research has been sponsored by the National Science Foundation through grants IIS-1652846, IIS-1423487, and ICER-1541043.

\bibliographystyle{splncs03}
\bibliography{./mlseb}

% %
% % ---- Bibliography ----
% %
% \begin{thebibliography}{}
% %
% \bibitem[1980]{2clar:eke}
% Clarke, F., Ekeland, I.:
% Nonlinear oscillations and
% boundary-value problems for Hamiltonian systems.
% Arch. Rat. Mech. Anal. 78, 315--333 (1982)
%
% \bibitem[1981]{2clar:eke:2}
% Clarke, F., Ekeland, I.:
% Solutions p\'{e}riodiques, du
% p\'{e}riode donn\'{e}e, des \'{e}quations hamiltoniennes.
% Note CRAS Paris 287, 1013--1015 (1978)
%
% \bibitem[1982]{2mich:tar}
% Michalek, R., Tarantello, G.:
% Subharmonic solutions with prescribed minimal
% period for nonautonomous Hamiltonian systems.
% J. Diff. Eq. 72, 28--55 (1988)
%
% \bibitem[1983]{2tar}
% Tarantello, G.:
% Subharmonic solutions for Hamiltonian
% systems via a $\bbbz_{p}$ pseudoindex theory.
% Annali di Matematica Pura (to appear)
%
% \bibitem[1985]{2rab}
% Rabinowitz, P.:
% On subharmonic solutions of a Hamiltonian system.
% Comm. Pure Appl. Math. 33, 609--633 (1980)
%
% \end{thebibliography}
% \clearpage
% \addtocmark[2]{Author Index} % additional numbered TOC entry
% \renewcommand{\indexname}{Author Index}
% \printindex
% \clearpage
% \addtocmark[2]{Subject Index} % additional numbered TOC entry
% \markboth{Subject Index}{Subject Index}
% \renewcommand{\indexname}{Subject Index}
% \input{subjidx.ind}
\end{document}

%% file: abstract.tex
Edge bundling methods can effectively alleviate visual clutter and reveal high-level graph structures in large graph visualization. Researchers have devoted
significant efforts to improve edge bundling according to different metrics. As the edge bundling family evolve rapidly, the \emph{quality} of edge bundles
receives increasing attention in the literature accordingly. In this paper, we present MLSEB, a novel method to generate edge bundles based on moving least
squares (MLS) approximation. In comparison with previous edge bundling methods, we argue that our MLSEB approach can generate better results based on a quantitative
metric of quality, and also ensure scalability and the efficiency for visualizing large graphs.
% However, It is challenging to ensure both the \emph{visual effect} and the \emph{quality} of bundling results.
%  

%% file: sec_intro.tex
\section{Introduction}
\label{sec:Introduction}

Traditional exploration methods of large graphs are often overwhelmed by severe visual clutter such as excessive vertex overlappings and edge crossings.
\emph{Edge bundling} is one of the effective approaches to reducing edge crossings in graph drawings. The main idea of edge bundling is to visually merge edges
with similar features (e.g., position, direction, and length) such that edge crossings are significantly reduced and the readability of graph drawings is
improved.

Substantial efforts have been made to develop various edge bundling algorithms to improve visual results. The current edge bundling family have provided a
diverse graph layouts that work with a wide spectrum of applications and domains based on different strategies or
metrics~\cite{Lhuillier:2017:SAE:3128397.3128448}. As the edge bundling techniques develop rapidly, the information visualization community is putting
increasing interests in evaluating the results of edge bundle drawings. The readability and faithfulness criteria are often used to evaluate graph drawings. Edge
bundling helps simplify graph drawings and increase readability, but yields distortion that makes it hard to preserve the faithfulness of original
graphs~\cite{DBLP:journals/corr/NguyenEH17}. To holistically address the evaluation of both readability and faithfulness for edge bundling visualization,
Lhuillier et al.~\cite{Lhuillier:2017:SAE:3128397.3128448} suggested a general metric where a ratio of clutter reduction to amount of distortion is computed
to measure the quality of edge bundling visualization. In this work, we aim to generate high-quality edge bundling results based on Lhuillier's suggestion, and
meanwhile ensure scalability and efficiency.

We introduce a novel edge bundling technique to generate edge bundles with moving least squares (MLS) approximation, namely MLSEB. Inspired by thinning an
unorganized point cloud to curve-like shapes~\cite{Lee:2000:CRU:342822.342830}, we use a distance-minimizing approximation function to generate bundle effects.
In particular, we first sample a graph into a point cloud data, and then use a moving least squares projection to generate curve-like bundles. Based on
Lhuillier's suggestion, we develop a quality assessment to evaluate edge bundling results. Using different real-world datasets, we demonstrate that MLSEB can
produce bundle results with a higher quality, and is scalable and efficient for large graphs by comparing different edge bundling methods. 

%% file: sec_related.tex
\section{Related Work}
\label{sec:Related Work}

% %------------------------------------------------
% \begin{figure*}[th!]
% \begin{center}
% $\begin{array}{c@{\hspace{0.01\linewidth}}c}
% \includegraphics[width=0.48\linewidth]{airline_fdeb} &
% \includegraphics[width=0.48\linewidth]{airline_mingle}
% \\
% \mbox{\small{(a) FDEB}} & \mbox{\small{(b) MINGLE }}
% \\
% \includegraphics[width=0.48\linewidth]{migration_gbeb} &
% \includegraphics[width=0.48\linewidth]{migration_wr}
% \\
% \mbox{\small{(c) GBEB}} & \mbox{\small{(d) WR}}
% \\
% \includegraphics[width=0.48\linewidth]{migration_sbeb} &
% \includegraphics[width=0.48\linewidth]{migration_kdeeb}
% \\
% \mbox{\small{(e) SDEB}} & \mbox{\small{(f) KDEEB}}
% \\
% \includegraphics[width=0.48\linewidth]{migration_cubu} &
% \includegraphics[width=0.48\linewidth]{migration_ffteb}
% \\
% \mbox{\small{(g) CUBu}} & \mbox{\small{(h) FFTEB}}
%
% \end{array}$
% \end{center}
% \vspace{-.1in}
% \caption[test caption]{Visualizing an airlines dataset with 2180
% edges (a-b) and a US migration dataset with 9780 edges (c-h) using the state-of-art
% methods.}
% %: (a) FDEB, (b) MINGLE, (c) GBEB, (d) WR, (e) SDEB, (f) KDEEB, (g) CUBu, and (h) FFTEB. }
% \vspace{-.2in}
% \label{fig:related_work}
% \end{figure*}
% %------------------------------------------------

%This section covers a wide variery of edge bundling methods. We note that
The recent study~\cite{Lhuillier:2017:SAE:3128397.3128448} has surveyed the state-of-the-art edge bundling techniques and their applications in a very detailed
manner. We revisit some of these methods by briefly summarizing the categories of the diverse bundling techniques. We consider our method as an image-based method, and hence we will discuss the 
image-based methods in more details. We will also cover some studies of quality evaluation in edge bundling and some studies on moving least squares approximation.

Holten~\cite{Holten:2006:HEB} pioneered the edge bundling techniques in graph drawings using a hierarchical structure.
%The work~\cite{4308636,Yu:2012:HSB:2311637.2311796} also use the similar idea.
\emph{Geometric-based} methods~\cite{DBLP:journals/tvcg/CuiZQWL08,5571244,Lambert:2010:WRR:2421836.2421848,DBLP:journals/tvcg/LuoLCM12} used a control mesh to guide bundling process.
\emph{Energy-based} minimization methods have been also used in many studies. Examples include ink-minimization
methods~\cite{conf/apvis/GansnerHNS11,Gansner:2006:ICL:1758612.1758651} and force-directed
methods~\cite{Holten09force,Nguyen2012,6065002,Zhou:2009:VCP:1925186,Zielasko:2016}. Most of these methods used compatibility criteria to measure the
similarity of different edges based on spatial information (i.e., length, position, angle, and visibility), and then moved the similar edges with ink-minimization or
force-directed strategies.

%\emph{Image-based} edge bundling techniques used density assessments to guide bundling process. A skeleton-based edge bundling was given in
%SBEB~\cite{Ersoy:2011:SEB:2068462.2068639}. Kernel density estimation edge bundling (KDEEB)~\cite{Hurter:2012:GBK:2322216.2322218} also built a skeleton for
%edge merging. The method first transformed an input graph into a density map using kernel density estimation. Sample points of edges moved towards the local
%density maxima to form bundles. Bottger et al.~\cite{DBLP:journals/tvcg/BottgerSLVM14} also used a similar advection method to generate bundles for visualizing
%the 3D correlations between neurons inside the human brain. Peysakhovich et al.~\cite{7156354} extended the basic idea of KDEEB and took attribute metrics into
%account to distinguish bundle based on different attributes. CUDA Universal Bundling (CUBu)~\cite{DBLP:journals/tvcg/ZwanCT16} used GPU parallelization to
%accelerate density assessment, and enabled bundling a graph with a million edges in interactive framerates. Moreover, Fast Fourier Transform Edge Bundling
%(FFTEB)~\cite{hurter2017PacificVis} addressed the scalability of density estimation by transforming the density space to the frequency space, thus speeding
%up the computation.

\emph{Image-based} techniques used a density assessment to guide bundling 
process~\cite{DBLP:journals/tvcg/BottgerSLVM14,Ersoy:2011:SEB:2068462.2068639,Hurter:2012:GBK:2322216.2322218,hurter2017PacificVis,7156354,DBLP:journals/tvcg/ZwanCT16}. These methods are generally 
based on \emph{Kernel Density Estimation}. Kernel density estimation edge bundling (KDEEB)~\cite{Hurter:2012:GBK:2322216.2322218} first transformed  an input graph into a density map using kernel 
density estimation, and then moved the sample points of edges towards the local density maxima to form bundles. Peysakhovich et al.~\cite{7156354} extended KDEEB using edge attributes to distinguish 
bundles. CUDA Universal Bundling (CUBu)~\cite{DBLP:journals/tvcg/ZwanCT16} used GPU acceleration to enable interactively bundling a graph with a million edges. Fast Fourier Transform Edge Bundling 
(FFTEB)~\cite{hurter2017PacificVis} improved the scalability of density estimation by transforming the density space to the frequency space.

% Figure~\ref{fig:related_work} (a) and (b) show the
% FDEB and MINGLE methods using an airlines dataset.
% Figure~\ref{fig:related_work} (c)-(h) show the results of GBEB, WR,
% SBEB, KDEEB, CUBu and FFTEB with an US migration dataset, respectively.

There are other edge bundling studies.
Bach et al.~\cite{7539373} investigated the connectivity of edge bundling methods on Confluent Drawings. Nguyen et al.~\cite{Nguyen:2012:SSE:2449763.2449801} proposed an edge bundling method
for streaming graphs, which extended the idea of TGI-EB~\cite{Nguyen2012}. Wu et al.~\cite{conf/bigdataconf/WuYY15} used textures to accelerate bundling for web-based
applications. Kwon et al.~\cite{7390081} showed their layout, rendering, and interaction methods for edge bundling in an immersive environment.

Serval studies introduced general metrics to quantify the
readability~\cite{dibattista1999graph,Purchase1997,Purchase1996,Tamassia:2007:HGD:1202383} and the faithfulness~\cite{Nguyen2013} of graph drawings. Some existing studies in edge bundling have defined quality assessments to evaluate the resulting
bundles. Nguyen et al.~\cite{DBLP:journals/corr/NguyenEH17} conducted a study on the faithfulness for force-directed edge bundling methods. Telea et
al.~\cite{DBLP:conf/dagstuhl/TeleaEHR09} posed a comparison between different hierarchical edge bundling methods. Telea et al.~\cite{telea_et_al:DSP:2009:2154}
surveyed the hierarchical edge bundling techniques and posed a comparison of the quality of bundled and unbundled graphs. Pupyrev et al.~\cite{Pupyrev2011} and
Kobourov et al.~\cite{Kobourov2014} worked towards measuring edge crossings. KDEEB~\cite{Hurter:2012:GBK:2322216.2322218} and
CUBu~\cite{DBLP:journals/tvcg/ZwanCT16} proposed post-relaxation if the distortion of edge bundles is too large, such that the mental map is preserved. For sequence graph edge bundling, 
Hurter et al.~\cite{6596126} used interpolation to preserve the mental map between sequence graphs. McGee et
al.~\cite{McGee:2012:ESI:2254556.2254670} conducted an empirical study on the impact of edge bundling.

Moving least squares (MLS) has been widely used to approximate smooth curves and surface from unorganized point
clouds~\cite{Alexa:2003:CRP:614289.614541,Levin2004,Mederos03movingleast}. Lee~\cite{Lee:2000:CRU:342822.342830} constructed a curve-like shape from unorganized
point clouds using an Euclidean minimum spanning tree. Least square projection (LSP) has been used in graph drawings~\cite{4378370}, where multidimensional data
points are projected into lower dimensions, while the similar relationship in neighboring points is preserved. 

%% file: sec_method.tex
\section{Background}

\subsection{Definition of Edge Bundling}

%~\cite{citeulike:5017658}, ~\cite{NME:NME1620370205} and ~\cite{Levin2003}

We first revisit a formal definition of edge bundling~\cite{hurter2017PacificVis}. Let $G = (V, E) \subset \R^{2}, V = \{v_i\} , E = \{e_i\}$ be a graph, where $v_i$ is a vertex and $e_i$ is an edge of $G$. Let $D: E \rightarrow
\R^{2}$ be a drawing operator, such that $D(G)$ represents the drawing of $G$ and $D(e_i)$ represents the drawing of an edge $e_i$. We define a
compatibility operator $\phi$, where $\phi(e_i, e_j)$ measures the similarity of two edges $e_i$ and $e_j$. Edges that are more similar than a threshold $\phi_{max}$ should
be bundled together, and $\phi$ can be used with some reasonable attributes and metrics(e.g., spatial information~\cite{Holten09force}). Let $B: \mathcal{D} \rightarrow \mathcal{D}$ be a bundling 
operation, where $\mathcal{D}\subset\R^{2}$ denotes the space of all graph drawings, and $B(D(e_i))$
denotes the resulting bundled drawing of $e_i$. For example, $D(e_i)$ can be a straight line drawing and $B(D(e_i))$ can be a drawing of curve or polyline. Hence, an edge
bundling algorithm can be expressed as:
%
%------------------------------------------------
\begin{equation}
\label{eq:equation_1}
\begin{split}
&\forall(e_i \in G, e_j \in G) | \phi(e_i, e_j) < \phi_{max} \rightarrow \\
&\delta(B(D(e_i)), B(D(e_j))) \ll \delta(D(e_i), D(e_j)),
\end{split}
\end{equation}
%------------------------------------------------
%
where $\delta$ is a distance metric in $\R^{2}$. Different edge bundling approaches explored various $\phi$, $B$, and $\delta$ to tackle
Equation~\ref{eq:equation_1} to gain different visual effects of edge bundling~\cite{Lhuillier:2017:SAE:3128397.3128448}.
% where $|\cdot|$ means the distance between two entities.

% %------------------------------------------------
% \begin{equation}
% \label{eq:equation_1}
% \begin{split}
% &\forall(e_i, e_j) \in E \times E | e_i \neq e_j \wedge \phi(e_i, e_j) < \phi_{max} \rightarrow \\
% &|B(D(e_i)), B(D(e_j)| \ll |D(e_i), D(e_j)|,
% \end{split}
% \end{equation}
% %------------------------------------------------

\subsection{Quality of Edge Bundling}

% The definition of faithfulness measures the information alteration or loss from
% edge bundling~\cite{Nguyen2013,DBLP:journals/corr/NguyenEH17}, i.e., how much
% information of an unbundled graph is conveyed in its corresponding bundled graph. The idea
% to preserve the faithfulness of a bundled graph is to make the distortion of the
% bundled graph as small as possible.

Edge bundling techniques trade the increase of readability for overdrawing by bending edges to form bundle effects.
Hence, edge bundle techniques naturally generate distortion from original graphs.
To quantify the quality of a bundled graph, Lhuillier et al.~\cite{Lhuillier:2017:SAE:3128397.3128448} suggested to use the ratio of \emph{clutter reduction} $C$ to \emph{amount of distortion} $T$ as a quality metric $Q$, i.e.,
%
%------------------------------------------------
\begin{equation}
\label{eq:equation_q}
{Q = \frac{C}{T}},
\end{equation}
%------------------------------------------------
%
In general, a larger $Q$ corresponds to a higher quality, and vice versa.
Lhuillier et al.~\cite{Lhuillier:2017:SAE:3128397.3128448} further posed a distortion measure. Simply, for an edge $e_i$, the distortion between an unbundled drawing $D(e_i)$ and a bundled result 
$B(D(e_i))$ is measured by computing the distance between them, i.e., $\delta(D(e_i), B(D(e_i)))$. Therefore, the overall distortion $T$ between an original unbundled graph and its bundled result can 
be defined as:
%
%------------------------------------------------
\begin{equation}
\label{eq:equation_4}
{T = \sum_{i=1}^n \delta(D(e_i), B(D(e_i)))},
\end{equation}
%------------------------------------------------
%
where $n$ is the number of edges. Equation~\ref{eq:equation_4} provides an intuitive metric to evaluate the distortion generated by a bundled graph.
%
%Equation~\ref{eq:equation_4} measures the overall distances between the unbundled and bundled graph.
%
The calculation of clutter reduction has not been fully concluded in the existing work. We propose a simple method to evaluate clutter reduction $C$, modify
Equation~\ref{eq:equation_4} to compute $T$, and then use $C$ and $T$ to quantify the quality $Q$ of edge bundling (Section~\ref{subsec:assess}).

% Hurter et al.~\cite{6596126,DBLP:journals/tvcg/HurterEFKT14} provided an example to demonstrate this metric: to preserve the faithfulness in dynamic bundled
% graphs, they interpolated the bundled layout using two consecutive frames, such that the dynamic graphs can transform smoothly, and the faithfulness of graphs
% can be preserved.

\section{Our Bundling Algorithm}
\label{sec:Method}

The main purpose of edge bundling is to achieve appealing bundle effects by bending edges, expressed by Equation~\ref{eq:equation_1}. Meanwhile, according to Equation~\ref{eq:equation_q}, an ideal 
algorithm should increase clutter reduction $C$, while decrease amount of distortion $T$, in order to achieve a higher quality $Q$ of edge bundling. Therefore, we should holistically address 
Equations~\ref{eq:equation_1} and~\ref{eq:equation_q}, which, however, has not been fully investigated in the existing work~\cite{Lhuillier:2017:SAE:3128397.3128448}.

\subsection{Sampling}

In general, given a graph $G$, a polyline is used to draw the line or curve presentation of an edge $e_i$. Sample points $x^i_k$, namely \emph{sites}, are used
to discretize the drawing of $e_i$. Formally,
%
%------------------------------------------------
\begin{equation}
\label{eq:equation_2}
{\{x^i_k | 1 \leq k \leq m_i\} \approx D(e_i)},
\end{equation}
%------------------------------------------------
%
where $m_i$ is the number of sites for $D(e_i)$. Note, many
methods~\cite{Hurter:2012:GBK:2322216.2322218,hurter2017PacificVis,7156354,DBLP:journals/tvcg/ZwanCT16} use a sampling step that is a small fraction of
the size of the display to sample each edge, which means the number of sites of $D(e_i)$ may be different. Similarly, the bundled drawing can also be
discretized as:
%
%------------------------------------------------
\begin{equation}
\label{eq:equation_extra2}
{B(\{x^i_k | 1 \leq k \leq m_i\}) \approx B(D(e_i))}.
\end{equation}
%------------------------------------------------
%
We measure the distortion between $D(e_i)$ and $B(D(e_i))$ by summing the Euclidean distance between each pair of $x^i_k$ and $B(x^i_k)$. Let $|\cdot|$ denote the Euclidean distance. Replace the edges in Equation~\ref{eq:equation_4}
using Equation~\ref{eq:equation_2} and Equation~\ref{eq:equation_extra2}, we have
%
%------------------------------------------------
\begin{equation}
\label{eq:equation_5}
{T = \sum_{i=1}^n(\sum_{k=1}^{m_i}|\{x^i_k\}, B(\{x^i_k\})|}).
\end{equation}
%------------------------------------------------
%
%
Similarly, Equation~\ref{eq:equation_1} can be modified as:
%
%------------------------------------------------
\begin{equation}
\begin{split}
\label{eq:equation_3}
&\forall(e_i \in G, e_j \in G) | \phi(e_i, e_j) < \phi_{max} \rightarrow \\
&|B(\{x^i_k\}), B(\{x^j_k\})| \ll |\{x^i_k\}, \{x^j_k\}|.
\end{split}
\end{equation}
%------------------------------------------------
%
% Similarly, by replacing the edges in Equation~\ref{eq:equation_4} using Equation~\ref{eq:equation_2}, we have
%
Therefore, we discretize each edge drawing $D(e_i)$ of $G$ by Equation~\ref{eq:equation_2}. All the sample points generated by Equation~\ref{eq:equation_2}
form a point cloud. According to Equation~\ref{eq:equation_3}, $x^i_k$ is moved to a new position $B(x^i_k)$ by a bundling operator $\emph{B}$. In the case of
kernel density estimation edge
bundling~\cite{Hurter:2012:GBK:2322216.2322218,hurter2017PacificVis,7156354,DBLP:journals/tvcg/ZwanCT16}, $x^i_k$ is moved to
$B(x^i_k)$ according to its local density gradient. These methods form the bundles by gathering sample points to their local density maxima, but do not consider the distortion of edges when moving
sample points. Therefore, certain artifacts, such as lattice effects and subsampled edge fragments, can be incurred. The methods, such as resampling and
post-relaxation~\cite{Hurter:2012:GBK:2322216.2322218,DBLP:journals/tvcg/ZwanCT16}, have been proposed to address these issues. However, these methods typically introduce a significant performance overhead that is challenging to alleviate~\cite{DBLP:journals/tvcg/ZwanCT16}.
We develop a new bundling operator $B$ with respect to Equation~\ref{eq:equation_3}, and minimize the distortion of each sample point locally. Moreover, our
method does not require resampling, and thereby can reduce the computational cost.

\subsection{Moving Least Squares Approximation}

%%------------------------------------------------
%\begin{figure*}[th!]
%\begin{center}
%$\begin{array}{ccc}
%\includegraphics[width=0.33\linewidth]{projection_x_i} &
%\includegraphics[width=0.33\linewidth]{projection_x_j} &
%\includegraphics[width=0.33\linewidth]{projection_all}
%\\
%%\mbox{\small{(a)}} & \mbox{\small{(b)}} & \mbox{\small{(c)}}
%\end{array}$
%\\
%\end{center}
%\vspace{-.1in}
%%\caption[test caption]{The sample (a), (c), (e), (g) and B-spline (b),
%%(d), (f), (h) results of 5 iterations using an US airlines example: iteration 0
%%(a, b); iteration 2 (c, d); iteration 5 (e, f); iteration 10 (g, h);}
%\caption[test caption]{The procedure of the MLS projection. The red and blue sites will move to new positions according to their projections on the implicit red curve and blue curve respectively. The implicit red curve and blue curve are defined by the red's and blue's neighborhood using MLS approximation respectively.}
%\
%\vspace{-.2in}
%\label{fig:projection}
%\end{figure*}
%%------------------------------------------------

We consider all the points formed by sampling, and assess the global distortion by expressing Equation~\ref{eq:equation_5} as:
%
%------------------------------------------------
\begin{equation}
\label{eq:equation_6}
\mathcal{T} = \sum_{i=1}^S{|x_i - B(x_i)|^2},
\end{equation}
%------------------------------------------------
%
where $x_i$ is a site in the point cloud, and $S$ is the number of sites of all edges. %We aim to minimize $\mathcal{T}$.

We assume there is a
\emph{skeleton} near $x_i$ and its neighborhood locally. A skeleton can be a
suitable place to gather curves to form
bundles~\cite{Ersoy:2011:SEB:2068462.2068639}. Assume a skeleton can be
interpreted as an implicit polynomial or piece-wise polynomial curve $f_i$,
which is unknown. The unknown $f_i$ can be gained by computing the coefficients
of $f_i$, i.e., by minimizing the following weighted least squares error $\epsilon$ within a set $\mathcal{H}(x_i)$ consisting of $x_i$ and its neighbor sites:
%
%------------------------------------------------
\begin{equation}
\label{eq:equation_7}
\epsilon = \sum_{j=1}^{h_i}{|x_j - f_i|^2\theta(|x_j - x_i|)},
\end{equation}
%------------------------------------------------
%
where $x_i \in \mathcal{H}(x_i)$, $x_j \in \mathcal{H}(x_i)$, $h_i$ is the size of $\mathcal{H}(x_i)$, and $|x_j - f_i|$ means the shortest Euclidean distance between $x_j$ and
$f_i$. We define the bundling operator $B$ on $x_i$ as a two-step procedure: first to construct $f_i$, and then to project $x_i$ onto $f_i$. The projected point is thereby $B(x_i)$ that is on $f_i$. The distance $|x_i - B(x_i)|$ from $x_i$ to
$B(x_i)$ is locally minimized by an appropriate nonnegative weighting function
$\theta$. The input of $\theta$ is $|x_j - x_i|$, which is the distance of
neighborhood $x_j$ to the site $x_i$. Instead of taking all sites of a graph into
account, we use a circle of radius $r$ (\emph{bandwidth}) centered at $x_i$ to collect the
neighborhood $x_j$ for $x_i$.

%------------------------------------------------
\begin{wrapfigure}{r}{0.3\linewidth}
\begin{center}
\includegraphics[width=0.8\linewidth]{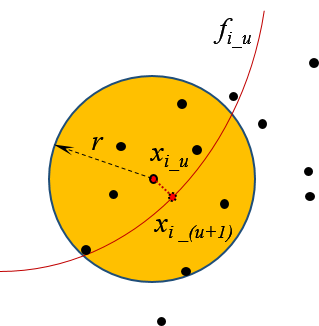}
\end{center}
\caption[test caption]{Two steps of our bundling operator $B$ on a site $x_{i\_u}$ in an iteration $u$. First, a local implicit regression curve $f_{i\_u}$ is constructed by the neighborhood of 
$x_{i\_u}$ with a bandwidth $r$ using the MLS approximation. Second, $x_{i\_u}$ is moved to a new position $x_{i\_(u+1)}$ that is the projection of $x_{i\_u}$ on $f_{i\_u}$.}
\
% \vspace{-.2in}
\label{fig:projection}
\end{wrapfigure}
%------------------------------------------------

If $\theta \equiv 1$, a least squares (LS) approximation is generated. However,
LS approximation does not work well to generate a polynomial curve that
locally reflects the density distribution of neighborhood. Alternatively, the moving least squares (MLS) method
can reduce a point cloud to a thin curve-like shape that is a near-best
approximation of the point set~\cite{Lee:2000:CRU:342822.342830,Levin2004}. Hence, we
use a local assessment to approximate $f_i$~\cite{citeulike:5017658}. The weighting
function we use is a cubic function~\cite{Mederos03movingleast}:
%
%------------------------------------------------
\begin{equation}
\label{eq:equation_8}
%\begin{aligned}
\theta(d) =
\begin{dcases*}
2\frac{d^3}{r^3} - 3\frac{d^2}{r^2} + 1 	& if $d < r$,
\\[3ex]
0 					 	& if $d \ge r$,
\end{dcases*}
%\end{aligned}
\end{equation}
%------------------------------------------------
%
where $d = |x_j - x_i|$. In this sense, minimizing Equation~\ref{eq:equation_7} leads to an MLS approximation so that $f_i$ is a local regression curve, and $|x_i - B(x_i)|$ is locally minimized. In 
other words, the distortion is locally minimized.

%Traditionally, a 2-step MLS projection procedure~\cite{Levin2004} has been used in many work to solve $f_i$~\cite{Alexa:2003:CRP:614289.614541,Mederos03movingleast}.

In our work, we use an MLS approximation to evaluate the distance $|x_j - f_i|$ for the neighborhood $\mathcal{H}(x_i)$ of $x_i$. Therefore, we use a basic projection~\cite{citeulike:5017658} to construct the implicit local regression curve $f_i$: We take a partial derivative of Equation~\ref{eq:equation_7} with respect to each coefficient of $f_i$, make each partial derivative equal to zero, and then solve the system of equations to generate all the coefficients of $f_i$~\cite{Nealen2004}.

\begin{figure*}[th!]
\begin{center}
$\begin{array}{c@{\hspace{0.01\linewidth}}c@{\hspace{0.01\linewidth}}c}
\includegraphics[width=0.33\linewidth]{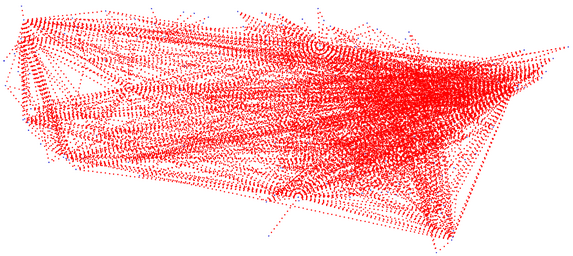} & \includegraphics[width=0.33\linewidth]{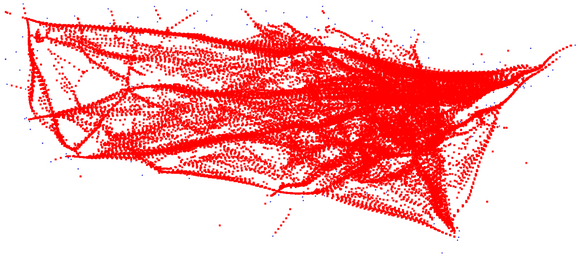} & \includegraphics[width=0.33\linewidth]{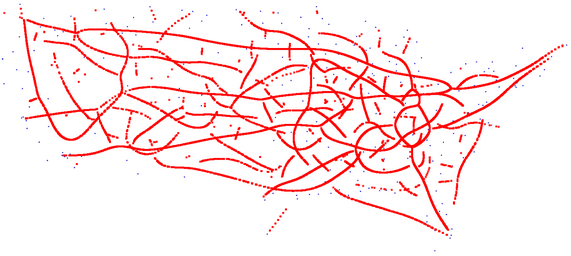} \\
\includegraphics[width=0.33\linewidth]{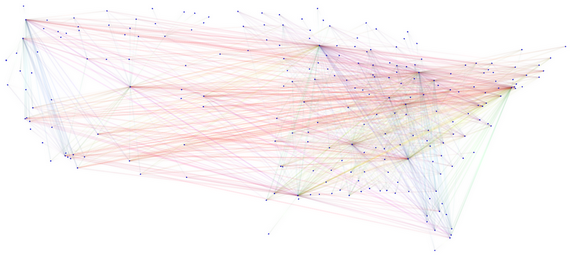} & \includegraphics[width=0.33\linewidth]{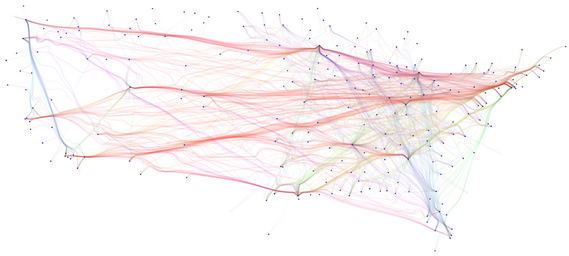} & \includegraphics[width=0.33\linewidth]{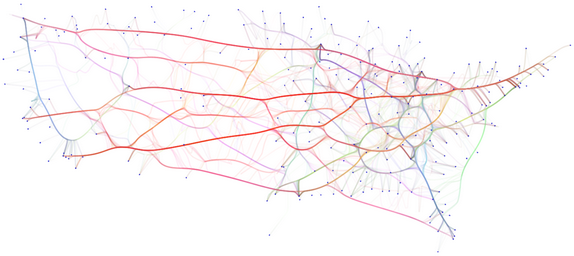} \\
\mbox{\scriptsize{Iteration 0}} & \mbox{\scriptsize{Iteration 2}} & \mbox{\scriptsize{Iteration 8}}
\end{array}$
\end{center}
\caption[test caption]{Using an US airlines dataset as an example, we first
sample each edge into a set of points (or sites). The resulting sites form a
point cloud (top-left). The top row shows the point cloud is converged
through an iterative MLS processing. The bottom row shows the corresponding
B-spline results. The first column shows the initial result before MLS. The
following columns show the results generated after the 2nd and 8th
iteration, respectively.}
\
% \vspace{-.2in}
\label{fig:iteration}
\end{figure*}

% However, one MLS projection is not sufficient to thin a point cloud to express a viable skeleton.
Similar to existing work~\cite{Ersoy:2011:SEB:2068462.2068639,Hurter:2012:GBK:2322216.2322218,hurter2017PacificVis,7156354,DBLP:journals/tvcg/ZwanCT16}, we
implement our bundling operator $B$ through an iteration strategy. In our method, two steps are applied iteratively, as shown in Figure~\ref{fig:projection}. We initially treat $x_i$ as $x_{i\_0}$. Then, in
each iteration $u$, the first step is to construct an optimal regression curve $f_{i\_u}$ by thinning the unordered point cloud within $\mathcal{H}(x_{i\_u})$,
the neighborhood of $x_{i\_u}$. In the second step, we project $x_{i\_u}$ onto $f_{i\_u}$ and obtain the projected point $x_{i\_(u+1)}$, i.e., $B(x_{i\_u})$.
In this way, a site $x_{i\_u}$ is moved to $x_{i\_(u+1)}$ based on the weighting function $\theta$ of its neighborhood $\mathcal{H}(x_{i\_u})$. Different from
the kernel density estimation
methods~\cite{Ersoy:2011:SEB:2068462.2068639,Hurter:2012:GBK:2322216.2322218,hurter2017PacificVis,7156354,DBLP:journals/tvcg/ZwanCT16}, MLS moves the site
$x_{i\_u}$ in the sense that the local error $\epsilon$ is bounded with the error of a local best polynomial approximation~\cite{Lee:2000:CRU:342822.342830}. In
our current work, this process stops when the iteration number reaches a predefined threshold. Then, for each edge, we compute a \emph{B-spline} curve based on
the final positions of its sites. Figure~\ref{fig:iteration} shows an example with two different iterations. For an illustration purpose, we show the
corresponding \emph{B-spline} curves for the iterations. In Figure~\ref{fig:iteration}, we can see that a curve-like skeleton is gradually formed from the point cloud
through the iterations in the top row, and a bundle effect becomes increasingly distinct as shown by the \emph{B-spline} results in the bottom row.

%------------------------------------------------

% site $x_i$ and its neighborhood define an implicit piecewise polynomial
% curve $f_i$ whose distances to $x_i$ and its neighborhood are holistically
% minimized,

Most of the existing image-based techniques use kernel density estimation (KDE), essentially, a mean-shift method that evaluates the local density maxima and
advects a site based on the gradients of the local density. However, KDE does not consider the distortion (Equation~\ref{eq:equation_4}) when moving
sample points, and thus resampling or post-relaxation is often required~\cite{Hurter:2012:GBK:2322216.2322218,DBLP:journals/tvcg/ZwanCT16}. Alternatively, our
MLSEB method uses an MLS approximation that projects a site $x_i$ to its local regression curve $f_i$, where $f_i$ is locally approximated by minimizing the distance
between $\mathcal{H}(x_i)$ and $f_i$ with a weighted function (Equation~\ref{eq:equation_7}). Therefore, the distance between its original position $x_i$ and its projected
position $B(x_i)$ is locally minimized based on the density of its neighborhood $\mathcal{H}(x_i)$. One advantage of our method is that MLS does not need to
resample each edge in bundling iterations because sites are projected into curves that do not generate over-converge artifacts or lattice effects. Fr\"{o}hlich
et al.~\cite{frohlich2014radial} showed that MLS produced better convergence results than KDE in biological studies. However, it remains an open question to
determine if KDE or MLS is better than one another in edge bundling. In Section~\ref{subsec:assess}, we will develop a quality assessment from
Equation~\ref{eq:equation_q}, and use it to evaluate and compare the quality of the drawings generated by our MLSEB method, the FFTEB method (a KDE-based method), and the FDEB method (a force-directed method). 

%% file: sec_implementation.tex
\section{Implementation}
\label{sec:Implementation}

% %------------------------------------------------
% \begin{figure*}[th!]
% \begin{center}
% $\begin{array}{|c|c|c}
% %{c@{\hspace{0.01\linewidth}}c@{\hspace{0.01\linewidth}}}
% \includegraphics[width=0.3\linewidth]{9grid_1} &
% \includegraphics[width=0.3\linewidth]{9grid_2} &
% \includegraphics[width=0.3\linewidth]{9grid_3}
% \\
% \mbox{\small{(a)}} & \mbox{\small{(b)}} & \mbox{\small{(c)}}
%
% \end{array}$
% \end{center}
% \vspace{-.1in}
% \caption[test caption]{The migration dataset (9780 edges): (a) FDEB (b)
% GBEB (c) WR (d) SDEB (e) KDEEB (f) CUBu (g) FFTEB (h) MLSEB}
% \vspace{-.2in}
% \label{fig:grid}
% \end{figure*}
% %------------------------------------------------

Our implementation involves simple data structures and computations, and thus
is easy to implement. First, we sample the edges of an input graph. We use the same
scheme as KDEEB's~\cite{Hurter:2012:GBK:2322216.2322218} to sample the input
edges with an uniform step $\rho$. The most time consuming step in our
method is gathering the neighborhood for every site. A typical solution in
a GPU implementation is to use \emph{Uniform Grid}~\cite{green2010particle} that subdivides the
space into uniformly sized cells. We use this method and set the size of the cell to be
$\frac{2}{3}r$ ($r$ is a prescribed radius or bandwidth) such that we can limit the
search space of each site to only cover at most 9 grid
cells~\cite{green2010particle}, thus avoid
a $O(S^2)$ search time for $S$ sites.

At the start of each iteration, all the sites are put into the corresponding
cells according to their current positions. This can be easily parallelized using
CUDA on a GPU~\cite{green2010particle}. Then, we project each site onto its
local regression line. The solution to compute the coefficients of
Equation~\ref{eq:equation_7} is introduced in the work~\cite{citeulike:5017658,Nealen2004}. It only requires a constant time to
solve the coefficients of a linear or quadratic system of equations. This can also be
parallelized using a GPU because computing the new projection position for every
site is independent.

To enhance the visualization of a bundled graph, we use the same shader scheme
of CUBu~\cite{DBLP:journals/tvcg/ZwanCT16}. We use the HSVA (i.e., hue $H$, saturation $S$, value $V$, and alpha $A$) color representation to visualize
edges. Each edge site $x_i$ is encoded with an HSVA value. We encode the direction and the length of the corresponding edge into $H$ and $S$, respectively. $V$ and $A$ are
used with a parabolic profile function $c(x) = \sqrt{1-2|t(x)-\frac{1}{2}|}$, and $t \in [0,1]$ is the edge arc-length parameterization. The functions of $V$ and $A$ are then $V(x) = 
\frac{l}{l_{max}}+(1-\frac{l}{l_{max}})c(x)$ and $A(x) = \alpha(1-\frac{l}{l_{max}}+\frac{l}{l_{max}}c(x))$ respectively, where $l$ is the length of the edge, $l_{max}$ is the longest edge in the 
graph, and $\alpha$ controls the overall transparency of all edges.
%
% %------------------------------------------------
% {\small
% \begin{equation}
% \label{eq:equation_9}
% \begin{array}{c@{\hspace{0.01\linewidth}}c@{\hspace{0.01\linewidth}}c}
% 
% \scriptsize{c(x) = \sqrt{1-2|t(x)-\frac{1}{2}|}}, & V(x) = \frac{l}{l_{max}}+(1-\frac{l}{l_{max}})c(x), & A(x) = \alpha(1-\frac{l}{l_{max}}+\frac{l}{l_{max}}c(x))
% 
% \end{array}
% \end{equation}
% }
% %------------------------------------------------
%
% where $l$ is the length of the edge, $l_{max}$ is the longest edge in the graph, $t \in [0,1]$
% is the edge arc-length parameterization, and $\alpha$ controls the overall transparency of all edges.

%------------------------------------------------
%%
%%------------------------------------------------
%\begin{equation}
%\label{eq:equation_10}
%V(x) = \frac{l}{l_{max}}+(1-\frac{l}{l_{max}})c(x),
%\end{equation}
%%------------------------------------------------
%%
%%------------------------------------------------
%\begin{equation}
%\label{eq:equation_11}
%A(x) = \alpha(1-\frac{l}{l_{max}}+\frac{l}{l_{max}}c(x)),
%\end{equation}
%%------------------------------------------------
%%

Next, we analyze the complexity of our MLSEB method. Similar to the existing KDE-based methods~\cite{Hurter:2012:GBK:2322216.2322218,hurter2017PacificVis,7156354,DBLP:journals/tvcg/ZwanCT16}, 
% MLSEB samples a graph into sites with a step $\rho$, and iteratively moves each site based on a weighted function of its neighborhood. 
MLSEB requires gathering neighbor sites for computation. After gathering, KDE-based methods conduct kernel splatting, gradient calculation, and site advection, which use a constant time for 
each site. In MLSEB, the time to solve Equation~\ref{eq:equation_7} and project a site to its local approximated curve is also constant for each site. Thereby, the complexity of MLSEB is the same as 
the traditional KDE-based methods, which is $O(I\cdot N \cdot S)$, where $I$ is the image resolution, $N$ is the number of bundling iterations, and $S$ is the number of sample points. However, MLSEB 
does not need additional operations, such as resampling, that are employed in the existing KDE-based methods.

We explore the parameter choices of MSLEB as follows. Similar to most the existing edge bundling methods, we use a step $\rho$, which is $5\%$ of the image resolution $I$, to sample each edge. The bandwidth, $r$, plays an important role in MLS to estimate the density information around each site. A
larger bandwidth captures more sample sites to reflect a more global feature, while a smaller bandwidth reveals a more local feature. By following a similar strategy in FDEB~\cite{Zielasko:2016} and KDEEB~\cite{Hurter:2012:GBK:2322216.2322218}, we decrease $r$ by a reduction factor $\lambda$ after each iteration.
Hurter et al.~\cite{Hurter:2012:GBK:2322216.2322218} stated that a kernel size follows an average density
estimation when $0.5 \leq \lambda \leq 0.9$. We set $r$ to be $5\% \leq r \leq 20\%$ of the display size $I$ to generate a stable edge-convergence
result. Through a heuristic study, we found that it is sufficient to yield good results by setting the iteration number $N$ between 3 and 10 and making the polynomial
order of $f_i$ in Equation~\ref{eq:equation_7} to be 1 or 2. 

%% file: sec_result.tex
\section{Results}
\label{sec:Result}

%------------------------------------------------------------
\input{./subsec_visual.tex}
%------------------------------------------------------------
\input{./subsec_assess.tex}
%------------------------------------------------------------

%% file: subsec_visual.tex
\subsection{Visualization and Performance Results}
\label{subsec:visual}

%------------------------------------------------
\begin{figure*}[th!]
\begin{center}
$\begin{array}{c@{\hspace{0.01\linewidth}}c}
\includegraphics[width=0.49\linewidth]{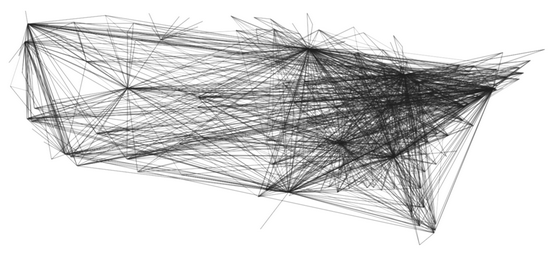} &
\includegraphics[width=0.49\linewidth]{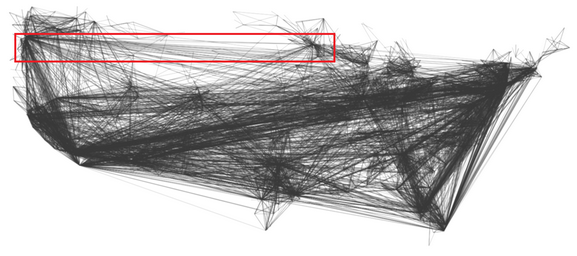}
\\
\mbox{\scriptsize{(a)}} & \mbox{\scriptsize{(b)}}
\end{array}$
\\
\centering{\mbox{\scriptsize{Original Node-link Diagrams}}}
\\
$\begin{array}{c@{\hspace{0.01\linewidth}}c}
\includegraphics[width=0.49\linewidth]{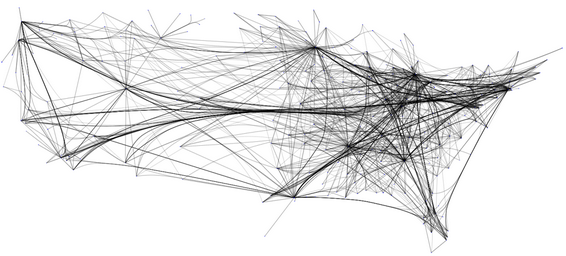} &
\includegraphics[width=0.49\linewidth]{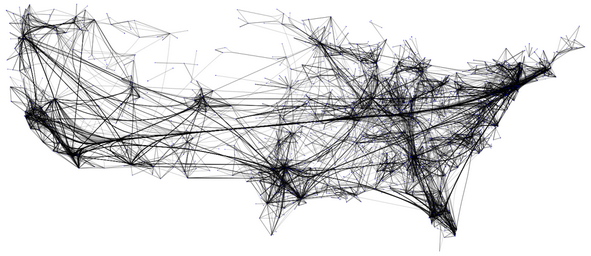}
\\
\mbox{\scriptsize{(c)}} & \mbox{\scriptsize{(d)}}
\end{array}$
\\
\centering{\mbox{\scriptsize{FDEB}}}
\\
$\begin{array}{c@{\hspace{0.01\linewidth}}c}
\includegraphics[width=0.49\linewidth]{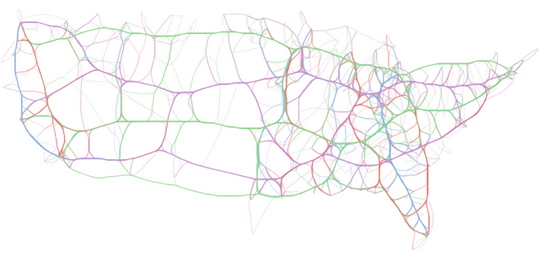} &
\includegraphics[width=0.49\linewidth]{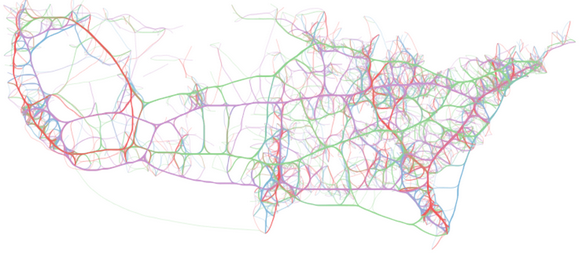}
\\
\mbox{\scriptsize{(e)}} & \mbox{\scriptsize{(f)}}
\end{array}$
\\
\centering{\mbox{\scriptsize{FFTEB}}}
\\
$\begin{array}{c@{\hspace{0.01\linewidth}}c}
\includegraphics[width=0.49\linewidth]{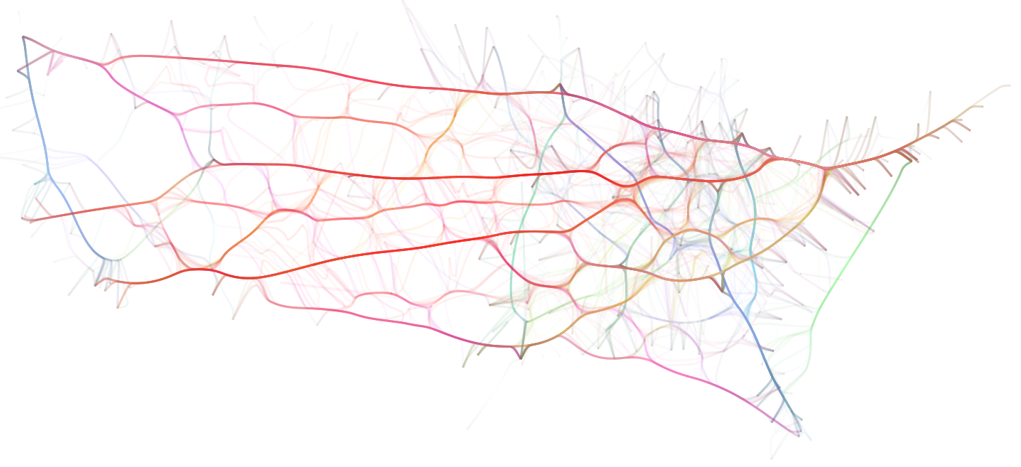} &
\includegraphics[width=0.49\linewidth]{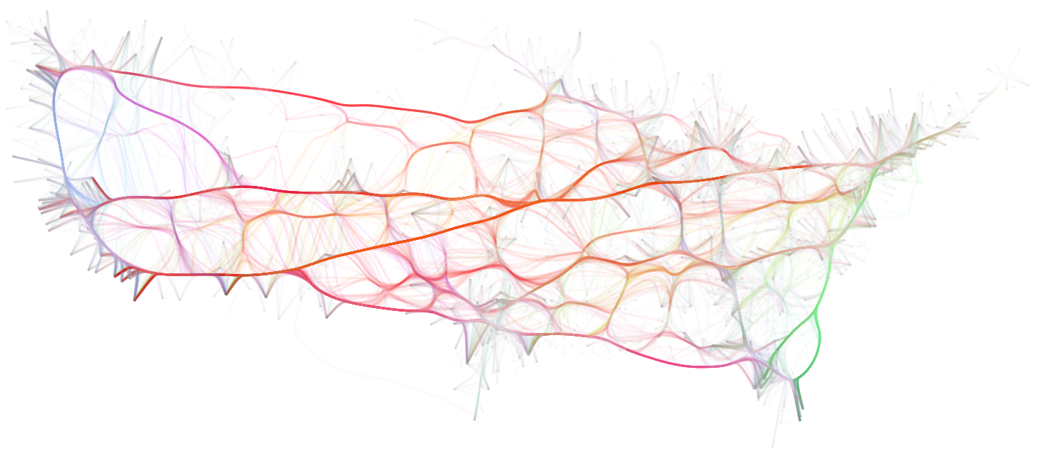}
\\
\mbox{\scriptsize{(g)}} & \mbox{\scriptsize{(h)}}
\end{array}$
\\
\centering{\mbox{\scriptsize{MLSEB}}}
\\
\end{center}
% \vspace{-.1in}
\caption[test caption]{Visualize the US airlines dataset (the left column) and the US
migrations dataset (the right column) with three different edge bundling methods, FDEB, FFTEB and MLSEB, respectively.}
\
% \vspace{-.2in}
\label{fig:result_1}
\end{figure*}
%------------------------------------------------

%------------------------------------------------
\begin{figure*}[th!]
\begin{center}
$\begin{array}{ccc}
\includegraphics[width=0.33\linewidth]{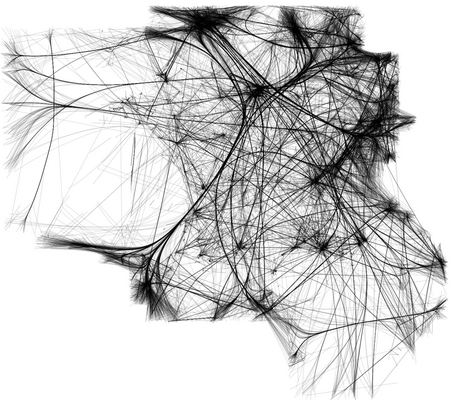} &
\includegraphics[width=0.33\linewidth]{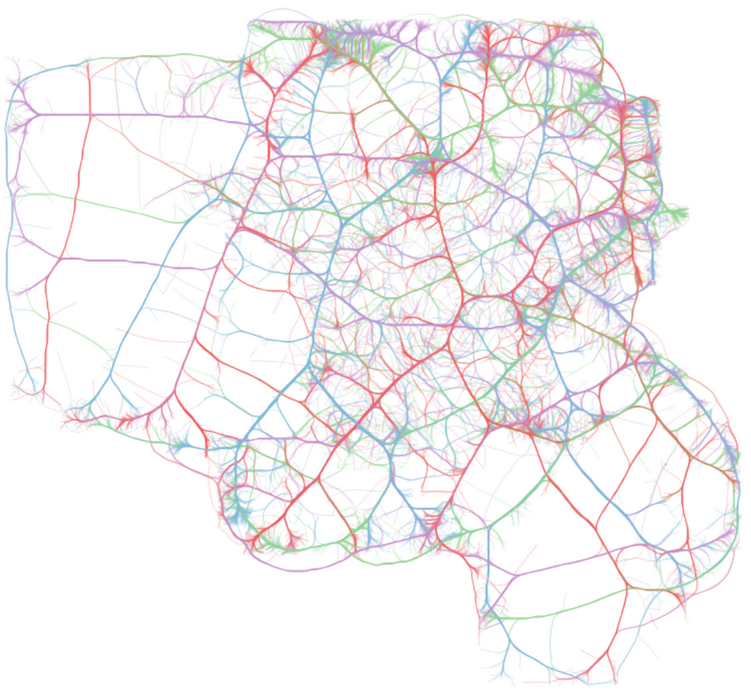} &
\includegraphics[width=0.33\linewidth]{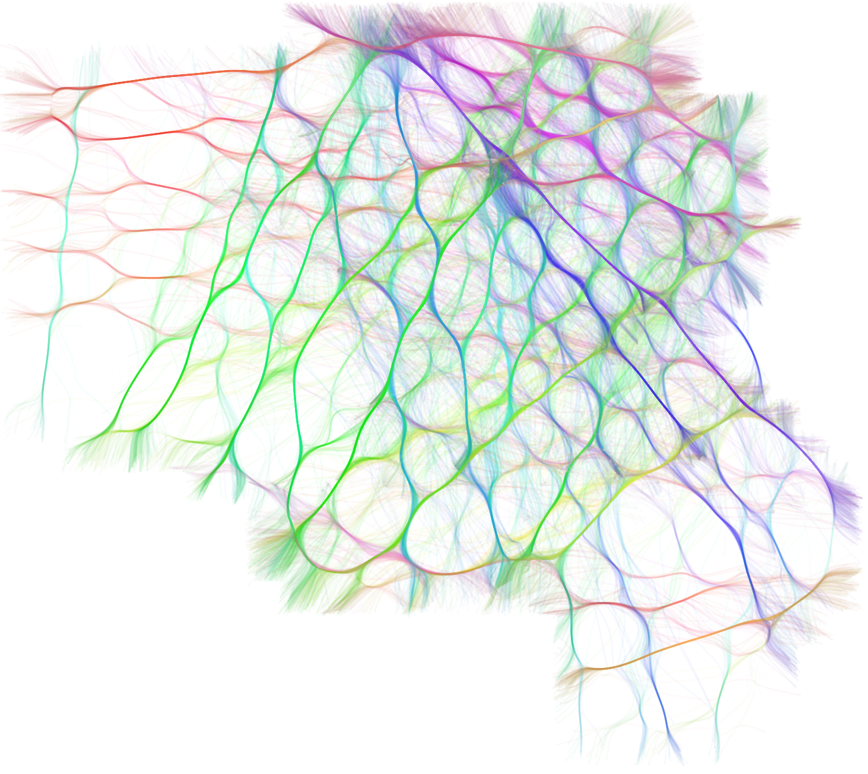}
\\
\mbox{\scriptsize{FDEB}} & \mbox{\scriptsize{FFTEB}} & \mbox{\scriptsize{MLSEB}}
\end{array}$
% \\
% \centering{\includegraphics[width=0.8\linewidth]{compare_france_mlseb}}
% \\
% \centering{\mbox{\small{MLSEB}}}
\end{center}
% \vspace{-.1in}
%\caption[test caption]{The sample (a), (c), (e), (g) and B-spline (b),
%(d), (f), (h) results of 5 iterations using an US airline example: iteration 0
%(a, b); iteration 2 (c, d); iteration 5 (e, f); iteration 10 (g, h);}
\caption[test caption]{Visualize the France airlines dataset (17274 edges) with
FDEB, FFTEB, and MLSEB.}
\
% \vspace{-.2in}
\label{fig:result_2}
\end{figure*}
%------------------------------------------------

%------------------------------------------------
\begin{figure*}[th!]
\begin{center}
$\begin{array}{c@{\hspace{0.01\linewidth}}c}
\includegraphics[width=0.49\linewidth]{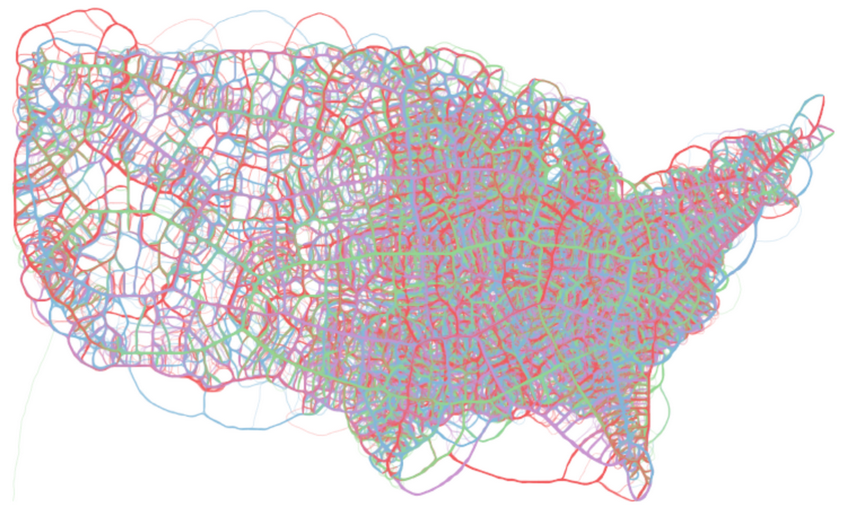} &
\includegraphics[width=0.49\linewidth]{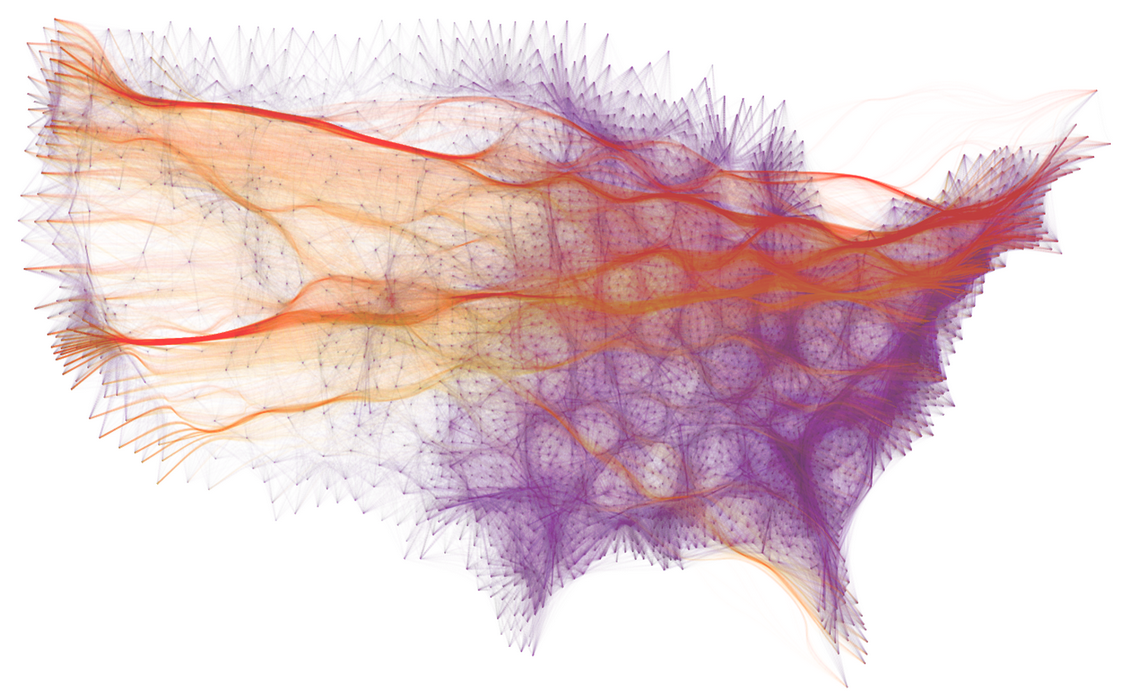}
\\
\mbox{\scriptsize{FFTEB}} & \mbox{\scriptsize{MLSEB}}
\end{array}$
\end{center}
% \vspace{-.1in}
\caption[test caption]{Comparison of FFTEB and MLSEB using a large US migrations dataset (545881 edges).}
\
% \vspace{-.2in}
\label{fig:performance}
\end{figure*}
%------------------------------------------------

We apply our MLSEB method to several graphs and compare its
effect and computational performance to the two existing
methods: FDEB that is the classic force-directed method, and FFTEB that is the latest enhanced KDE-based method of image-based edge bundling algorithms (such as KDEEB and CUBu).

The left column in Figure~\ref{fig:result_1} compares the visualization results of our MLSEB
method with other bundling methods using the US airlines dataset (2101 edges). Our MLSEB method provides similar results, and generates tight, smooth and locally well-separated bundles. High-level
graph
structures are also revealed in our results. The right column in Figure~\ref{fig:result_1}
shows the comparison using the US migrations dataset (9780 edges). Figure~\ref{fig:result_2} shows another example using the France
airlines dataset with 17274 edges. In these results, the
main migration and airline patterns are clearly revealed using MLSEB. In the migrations dataset, FDEB and FFTEB fall short in showing some subtle structures of the original
graph. For example, in the original node-link diagram of Figure~\ref{fig:result_1}(b), the edges (within the red box) connect the city of Portland to some cities in the northern U.S are distorted 
significantly from their original positions in the results of the FDEB (Figure~\ref{fig:result_1}(d)) and FFTEB (Figure~\ref{fig:result_1}(f)), while our MLSEB result has a distinguished bundle 
effect that reveals this subtle graph structure. In
Figure~\ref{fig:performance}, we compare the visual result of MLSEB to FFTEB using a large US migrations dataset with 545881 edges. We encode the color of a edge with only its length in this
example. MLSEB shows more long-length edge patterns than FFTEB.

Table~\ref{table:performance} shows the performance comparison between our
MLSEB method and the current fastest edge bundling method FFTEB. In our
performance comparison, we used the US airlines graph, the US migrations graph, the France airlines graph, and the large US migrations graph. The timing results for MLSEB and FFTEB are based
on one iteration, and we excluded the timing of memory allocation and data transferring for
both methods. The devices used in our experiments are a desktop with an 8X
Intel Core i7-6700K 4.0GHz CPU with 32GB memory and a NVIDIA GeForce GTX TITAN X
GPU.
Comparing with the fastest algorithm FFTEB in the state-of-the-art, we can clearly see
that MLSEB is at the same order of magnitude of FFTEB in terms of computational
speed, as shown in Table~\ref{table:performance}.

%----------------------------------
\begin{table}[t!]
\caption{Performance comparison.}
\begin{center}
\scalebox{0.8}{
% \resizebox{\textwidth}{!}{
\begin{tabular}{ l | r | r | r | r | r }
\toprule
\textbf{Graph} & \textbf{Edges} & \multicolumn{2}{c|}{\textbf{FFTEB}} &  \multicolumn{2}{c}{\textbf{MLSEB}} \\
 & & \makebox[2cm][r]{Samples} & \makebox[2cm][r]{Time (ms)} & \makebox[2cm][r]{Samples} & \makebox[2cm][r]{Time (ms)} \\ \midrule
US airlines     		& 2180 	& 105K  & 40 & 85K & 22 \\
US migrations   	& 9780 	& 489K  & 48 & 207K & 38 \\
France airlines 		& 17274 & 864K  & 70 & 990K & 94 \\
Large US migrations 	& 545881 & 6.4M  & 123 & 5.8M & 554 \\
% \cellcolor[gray]{0.8}
\bottomrule
\end{tabular}
}
% }
\end{center}
\label{table:performance}
\end{table}
%----------------------------------

%%------------------------------------------------
%\begin{figure*}[th!]
%\begin{center}
%\includegraphics[width=0.8\linewidth]{bundling_time}
%% \includegraphics[width=0.8\linewidth]{example_2}
%\\
%%\mbox{\small{Timing results with the US airline, the US migration and the France airline dataset}}
%% \\
%% \centering{\includegraphics[width=0.8\linewidth]{compare_france_mlseb}}
%% \\
%% \centering{\mbox{\small{MLSEB}}}
%\end{center}
%% \vspace{-.1in}
%\caption[test caption]{Comparison of timing results of FFTEB and MLSEB as a function of the number of sampling sites.}
%\
%% \vspace{-.2in}
%\label{fig:performance}
%\end{figure*}
%%------------------------------------------------

%We also compare the scalability of FFTEB and MLSEB in Figure~\ref{fig:performance}. It shows the computing time of MLSEB is almost linear with the size of sampling sites.

%Our comparison is not only focus on performance result. Quality assessment is discussed next. In Section~\ref{subsec:assess}, we will first discuss a quality metric to evaluate the quality of produced bundling drawings, and we will compare our result with other methods based on the quality metric.

%% file: subsec_assess.tex
\subsection{Quality Assessment of Bundled Graphs}
\label{subsec:assess}

Apart from comparing the visualization and performance results, we propose a quality metric to evaluate the quality of bundling drawings based on Equation~\ref{eq:equation_q}.

Equation~\ref{eq:equation_q} gives a general quality metric $Q$ based on the ratio of clutter reduction $C$ to amount of distortion $T$. However, the
quantification of clutter reduction $C$ has been not fully concluded in existing work.
%Inspired by Hurter et al.~\cite{Lhuillier:2017:SAE:3128397.3128448},
We propose to employ the reduction of the used pixel number $\Delta{P}$ in a graph drawing to measure $C$. Specifically, $C = \Delta{P} = P - P^{\prime}$ that
is the difference of the used pixel number $P$ of the original drawing and the used pixel number $P^{\prime}$ of the bundled drawing.
%%
%%------------------------------------------------
%\begin{equation}
%\label{eq:equation_14}
%\Delta{P} = P - P^{\prime}.
%\end{equation}
%%------------------------------------------------
%%

Intuitively, $T$ can be given by Equation~\ref{eq:equation_5} that quantifies the total distortion of all the sample points. However, different methods can generate different numbers of sample points. For example, FDEB generates the same number of sample points for each edge, while our MLSEB method and the KDE-based methods sample different edges into different numbers of points. Thus, instead of the total distortion of all the sample points, we use the average distortion: $\overline{T} = \frac{T}{S}$,
%%
%%------------------------------------------------
%\begin{equation}
%\label{eq:equation_13}
%\overline{T} = \frac{T}{S},
%\end{equation}
%%------------------------------------------------
%%
where $S$ is the total number of the sample points in the graph. Therefore, we modify Equation~\ref{eq:equation_q} to
%
%------------------------------------------------
\begin{equation}
\label{eq:equation_12}
Q = \frac{\Delta{P}}{\overline{T}}.
\end{equation}
%------------------------------------------------
%
%can be
%largely different, which is hard to evaluate the distortion without biases.
%
%a general quality assessment $Q$:
%%
%%------------------------------------------------
%\begin{equation}
%\label{eq:equation_12}
%Q = \frac{\Delta{P}}{\overline{D}},
%\end{equation}
%%------------------------------------------------
%%
%where $\overline{D}$ is the quantitative information of the average distortion
%of each site in bundled graph, i.e., how many pixels each
%sampling site moves in average. Specifically,
%%
%%------------------------------------------------
%\begin{equation}
%\label{eq:equation_13}
%\overline{D} = \frac{D}{n},
%\end{equation}
%%------------------------------------------------
%%
%where $n$ is the total number of the sampling sites in the graph. The reason why we
%use the average distortion of each site instead of the total amount of
%distortion is because the number of sampling sites in different methods can be
%largely different, which is hard to evaluate the distortion without biases. We
%can normalize the distortion by using the average distortion of each
%site.
%
%
%
The rationale of Equation~\ref{eq:equation_12} is to measure how many pixels are decreased
by generating one unit distortion. A higher value of $Q$ means a better quality result.
Table~\ref{table:quality_assessment} shows the quantitative quality comparison between our MLSEB method, FDEB and FFTEB. Our comparison is based on the drawings with an image
resolution of $400 \times 400$, as shown in Figures~\ref{fig:result_1},~\ref{fig:result_2}, and~\ref{fig:performance}. All the
statistic results are generated after a
graph is bundled, i.e., after all iterations. We note that it makes less
sense to compare the distortion in each iteration because the initial iterations
of some methods, such as FDEB and FFTEB, may have surprisingly large distortion.
It is more reasonable to compare the quality of results after the bundling
iterations are finished. We also note that using different parameters, such as
different iteration numbers and different bandwidths %(image-based methods and MLSEB)
for different methods, can yield different results. We use the recommended parameters
in FDEB's and FFTEB's papers~\cite{Holten09force,hurter2017PacificVis}, which
are the best results we can get from the existing work. The $S$ columns in
Table~\ref{table:quality_assessment} show the numbers of the sample points in a graph using different methods.

%----------------------------------
\begin{table}[t!]
\caption{Quality comparison using the US migrations graph.}
\begin{center}
\resizebox{\textwidth}{!}{
\begin{tabular}{l|r | rrrrr | rrrrr | rrrrr}
\toprule
\textbf{Graph} & \textbf{Edges} & \multicolumn{5}{c|}{\textbf{FDEB}} &  \multicolumn{5}{c|}{\textbf{FFTEB}} & \multicolumn{5}{c}{\textbf{MLSEB}} \\ %\cmidrule{3-8} \cmidrule{9-14} \cmidrule{15-20}
%  & & \multicolumn{2}{l}{FP}  & \multicolumn{2}{l}{GCV} & \multicolumn{2}{l}{REML} & \multicolumn{2}{l}{FP} & \multicolumn{2}{l}{GCV} & \multicolumn{2}{l}{REML}
% & \multicolumn{2}{l}{FP} & AIC & REML \\
 & & \makebox[1cm][r]{$S$} & \makebox[1cm][r]{$P$} & \makebox[1cm][r]{$P^\prime$} & \makebox[1cm][r]{$\overline{T}$} & \makebox[1cm][r]{$Q$} & \makebox[1cm][r]{$S$} & \makebox[1cm][r]{$P$} & \makebox[1cm][r]{$P^\prime$} & \makebox[1cm][r]{$\overline{T}$} & \makebox[1cm][r]{$Q$} & \makebox[1cm][r]{$S$} & \makebox[1cm][r]{$P$} & \makebox[1cm][r]{$P^\prime$} & \makebox[1cm][r]{$\overline{T}$} & \makebox[1cm][r]{$Q$} \\ \midrule
% US airlines     & 2180 	& 91k  & 32k & 19k & 0.88k & \cellcolor[gray]{0.8}14.4   & 105k & 32k & 18k & 1.2k & \cellcolor[gray]{0.8}11.9 	 	 & 813k  & 32k & 25k & 1.10  & \cellcolor[gray]{0.8} 6.2\\
% US migrations   & 9780 	& 96k  & 33k & 25k & 0.92k & \cellcolor[gray]{0.8}9.20   & 489k & 32k & 24k & 1.0k & \cellcolor[gray]{0.8}7.60 		 & 3785k & 34k & 26k & 0.88  & \cellcolor[gray]{0.8} 8.9\\
% France airline & 17274 	& 161k & 81k & 60k & 0.80k & \cellcolor[gray]{0.8}26.0   & 864k & 81k & 57k & 1.6k & \cellcolor[gray]{0.8}21.3 		 & 6685k & 81k & 72k & 2.6   & \cellcolor[gray]{0.8} 3.7\\

US airlines     & 2180 	& 813K  & 32K & 25K & 1.10K  & \cellcolor[gray]{0.8} 6.2		 & 105K & 32K & 18K & 1.2K & \cellcolor[gray]{0.8} 11.9 	
	 & 85K  & 32K & 19K & 0.88K & \cellcolor[gray]{0.8} 14.4 \\
US migrations   & 9780 	& 3785K & 34K & 26K & 0.88K  & \cellcolor[gray]{0.8} 8.9		 & 489K & 32K & 24K & 1.0K & \cellcolor[gray]{0.8} 7.60 	
	 & 207k  & 33k & 25k & 0.92k & \cellcolor[gray]{0.8} 9.20	\\
France airlines & 17274 	& 6685K & 81K & 72K & 2.60K   & \cellcolor[gray]{0.8} 3.7	 	 & 864K & 81K & 57K & 1.6K & \cellcolor[gray]{0.8} 21.3 	
	 & 990K & 81K & 60K & 0.80K & \cellcolor[gray]{0.8} 26.0 \\
Large US migrations & 545881 	& n/a & n/a & n/a & n/a   & n/a	 	 & 6.4M & 108k & 84k & 1.8k & \cellcolor[gray]{0.8} 13.3
	
	 & 5.8M & 107k & 95k & 0.90 & \cellcolor[gray]{0.8} 13.3 \\
\bottomrule
% \cellcolor[gray]{0.8}
\end{tabular}}
\end{center}
\label{table:quality_assessment}
\end{table}
%----------------------------------

We can see that the quality of MLSEB is generally better than the other two methods in
terms of Equation~\ref{eq:equation_12}. For the four different datasets, FFTEB
makes the most clutter reduction. However, it also incurs more distortion. FDEB achieves a comparable quality as ours for the US migrations dataset;  whereas, when
the dataset is getting larger (France airlines), FDEB will generate tremendous
distortion, as shown in Table~\ref{table:quality_assessment} and
Figure~\ref{fig:result_2}, thus lowering the quality score. Note when using the large US migrations dataset, the advantage of MLSEB over FFTEB becomes marginal. Overall, MLSEB gains the
highest quantitative scores in terms of quality according to Equation~\ref{eq:equation_12}.
% Overall, MLSEB generates visually appealing
% results for the four different datasets, and at the same time, gains the
% highest quantitative scores in terms of quality according to
% Equation~\ref{eq:equation_12}. 

%% file: sec_conclusion.tex
\section{Conclusions and Future Work}
\label{sec:Conclusion}

We present a new edge bundling method MLSEB that holistically considers distortion minimization and clutter reduction. Inspired by
the MLS work~\cite{Alexa:2003:CRP:614289.614541,Lee:2000:CRU:342822.342830},
our approach generate bundle effects by iteratively projecting each site to its local regression curve to converge with
other nearby sites based on its neighborhood's density. 
Such a local regression curve can reduce the distortion of the local bundle.
Our method is easy to implement. The timing result shows MLSEB is at the same order of magnitude of the current fastest edge bundling method FFTEB in terms of computational speed.

We use a quality assessment to evaluate the quality of resulting edge bundles. Our MLSEB method shows better results in our preliminary comparison. However, a more comprehensive comparison between our MLSEB method and the other methods requires further investigation, where other factors (e.g., edge crossing reduction) may be also considered.
In addition, we plan to apply optimal bandwidth selection~\cite{Lipman:2006:EBO:1281957.1281966,4745637} to improve MLSEB.
%The challenge of automatic parameter selection was also posed~\cite{Lhuillier:2017:SAE:3128397.3128448}.
%Another limitation of MLSEB is that it has not taken
%edge attributes into the MLS approximation. 
%
%Although our visualization can
%reveal some basic attributes of edges (such as, length and direction) using shader
%functions, our MLS approximation does not couple the semantic attributes with the
%approximation. 
We would also like to incorporate semantic attributes into MLSEB to enhance bundling results.
Last but not least, bundling a very large graph (e.g., one with billions or trillions of edges) remains
a very challenging task, which is a next possible direction in our future work.